# Transparency guided ensemble convolutional neural networks for stratification of pseudoprogression and true progression of glioblastoma multiform

Xiaoming Liu[b], Michael D. Chan[d], Xiaobo Zhou[c], Xiaohua Qian[a]

[a] Institute for Medical Imaging Technology, School of Biomedical Engineering, Shanghai Jiao Tong University, Shanghai 200030, China

[b] College of electronic science and engineering, Jilin University

[c] School of Biomedical Informatics, The University of Texas Health Science Center at Houston, Houston, TX 77030, USA

[d] Department of Radiology, Wake Forest School of Medicine, Winston-Salem, NC 27157, USA

*ABSTRACT*

Pseudoprogression (PsP) is an imitation of true tumor progression (TTP) in patients with glioblastoma multiform (GBM). Differentiating them is a challenging and time-consuming task for radiologists. Although deep neural networks can automatically diagnose PsP and TTP, interpretability shortage is always the Achilles's heel. To overcome these shortcomings and win physician's trust, we propose a transparency guided ensemble convolutional neural network to automatically stratify PsP and TTP on magnetic resonance imaging (MRI). A total of 84 patients with GBM are enrolled in the study. First, three typical convolutional neutral networks (CNNs)- VGG, ResNet and DenseNet- are trained to distinguish PsP and TTP on the dataset. Subsequently, we use the class-specific gradient information from convolutional layers to highlight the important regions in MRI. Radiological experts are then recruited to select the most lesion-relevant layer of each CNN. Finally, the selected layers are utilized to guide the construction of multi-scale ensemble CNN. The classified accuracy of the presented network is 90.20%, the promotion of specificity reaches more than 20%. The results demonstrate that network transparency and ensemble can enhance the reliability and accuracy of CNNs. The presented network is promising for the diagnosis of PsP and TTP.

## 1. Introduction

Glioblastoma multiforme (GBM) is the most common malignant central nervous system tumor in adults.[1] Currently, the standard treatment for GBM is surgical resection followed by radiation therapy with concurrent temozolomide (TMZ) and adjuvant TMZ. Despite of the benefits for patients, such treatment faces a new dilemma: the incidence of pseudoprogression (PsP). PsP is caused by post-treatment reactions, such as inflammation, ischemia or radiation necrosis [2] and will occur in about a third of all patients with GBM. PsP is a contrast enhancement that mimics early tumor progression, but unlike true tumor progression (TTP), it can improve or stabilize spontaneously without intervention.[3]

Accurate diagnosis of PsP and TTP is critical because the results may directly influence the therapy strategy option and overall survival of patients. Although brain tumor biopsies can effectively differentiate PsP, the invasive procedure and second surgery may cause more risks of patients. Besides, follow-up imaging can provide an accurate identification, but it may discourage the patients of TTP to capture the best treatment opportunity. Diffusion tensor imaging (DTI) is a type of magnetic resonance imaging (MRI) which is good at anisotropy measurement. Since brain tissues with PsP have lower FA values than TTP, DTI is regarded as a potential way to truly reflect

PsP and TTP.[4,5,6] However, because PsP is very similar with TTP in not only intensity but shapes, it is still difficult for radiologists identifying them by DTI in the clinic. Therefore, a fast and non-invasive method is extremely needed to diagnose PsP and TTP.

Techniques of computer-aided diagnosis (CAD) can fully take use of the gray information in medical images and then extract huge amount of features. The combination of minable features and advanced algorithms is more likely to outperform human in some medical diagnosis cases. Several attempts of CAD, such as parametric response maps [7,8,9], gray-level co-occurrence matrix [10] and dictionary learning [6] are applied in the diagnosis of PsP and TTP. However, these methods cannot utilize high-dimensional features and thus fail to capture subtle differences between PsP and TTP in MRIs. Overall, the performances on diagnosis system of PsP and TTP are expected to be improved.

Convolutional neutral networks (CNNs) are the emerging technique for image classification. CNNs are composed of many filter-based layers and can learn high-level features of the input images. CNNs consist of three types of layers: convolutional, pooling and fully connected. For each convolutional layer $l$, the input image is firstly convolved with several convolutional kernels $W$ and then corresponding bias b is added. If the number of the kernel is $K$, the feature map $X_k$ of $l$ can be computed as: $X_k^l = \sigma(W_k^{l-1} \cdot X^{l-1} + b_k^{l-1})$. $\sigma(\cdot)$ is the activation function which performs non-linear transformation of $X_k$. Generally, convolutional layers are followed by pooling layers. Pooling layers aim at compressing the features. The typical pooling operations include average pooling and max pooling. Besides, fully-connected layers are commonly employed at the end of the networks and act as the classification layers. After the breakthrough of so-called AlexNet [11] -one of the most typical CNNs- a series of improved CNNs are appeared. For example, VGG [12] and GoogleNet [13] can realize the deeper and wider architectures respectively. Additionally, ResNet [14] and DenseNet [15] show outstanding performances via feature reorganization.

CNNs makes many achievements in the medical field. And they are also applied to the diagnosis and prognosis of nervous-system diseases, such as the Alzheimer's diagnostics [16,17], brain tumor grading [18] and the survival prediction of brain tumor patients [19,20]. Although CNNs have shown superior performances in various tasks, the black-box representations make it difficult for human to understand the abstract features inside the networks. In order to understand how and why CNNs work, there are already some works trying to interpret the networks. Paper [21] [22] visualize final predictions of CNN by directly highlighting pixels in high probability. In addition, some gradient-based methods [23,24] use the gradients to estimate the feature appearance. Compared with the other applications, visualization and explanation are especially important in medical CNNs. General speaking, interpretability can contribute in three aspects: (i) It can build up physician's trust in intelligent systems and develop the clinical use of CAD systems. (ii) It can make researchers understand the networks better. That is to say we are likely to promote the CNNs with targets for different problems. (iii) It can visualize the high-dimensional features in CNNs comprehensible features can speed up the intelligence combination between human and computer and thus may help doctors to discover some new imaging biomarker of diseases. Therefore, interpretable CNNs can make feedbacks to medical researches and provide new guidance in disease treatment.

In this paper, we present transparency guided multi-scale ensemble CNN to promote deep feature understanding and reuse and differentiate PsP and TTP more accurately and reliably. As shown in Fig.1, first, three classification CNNs including VGG, RestNet and DenseNet are trained on DTI dataset separately. Next, we try to make the three models more transparent by visualizing the 'importance' of all the feature maps layer by layer. We find that totally different views are focused between layers even in the same network. For example, some layers pay more attention to edges while others highlight the differences between pixels. Therefore, we then invite radiologists to give experience-driven explanation of the visualization. They are asked to locate the layers which are correlated most with the lesion regions in three CNNs respectively. To incorporate multiple features from different networks, we finally build an ensemble network using the feature maps of the selected layers. A multi-scale processing strategy is also introduced in the presented ensemble network with the aim of multi-perspective feature complementation.

Overall, the research is novel with the following three contributions. First, the presented framework dedicates to explain and understand the black-box CNNs. Additionally, the human knowledge and experience are introduced in the feature selection. It can enhance reliability as well as the communications between doctors and computers. Second, an ensemble CNN is established in this research. The ensemble CNN fuses multiple features from different architectures. In this way, it can compensate shortcomings of single CNNs and promote accuracy of classification. Finally, the study proposes a multi-scale network. The model can use features in various perspectives and thus expands local information to global information.

The details of material are given in Section 2, related works are review in Section 3, the presented ensemble network is described in Section 4, the process of experiment and results are displayed in Section 5, some discussion are given in Section 6. Finally, we conclude in Section 7.

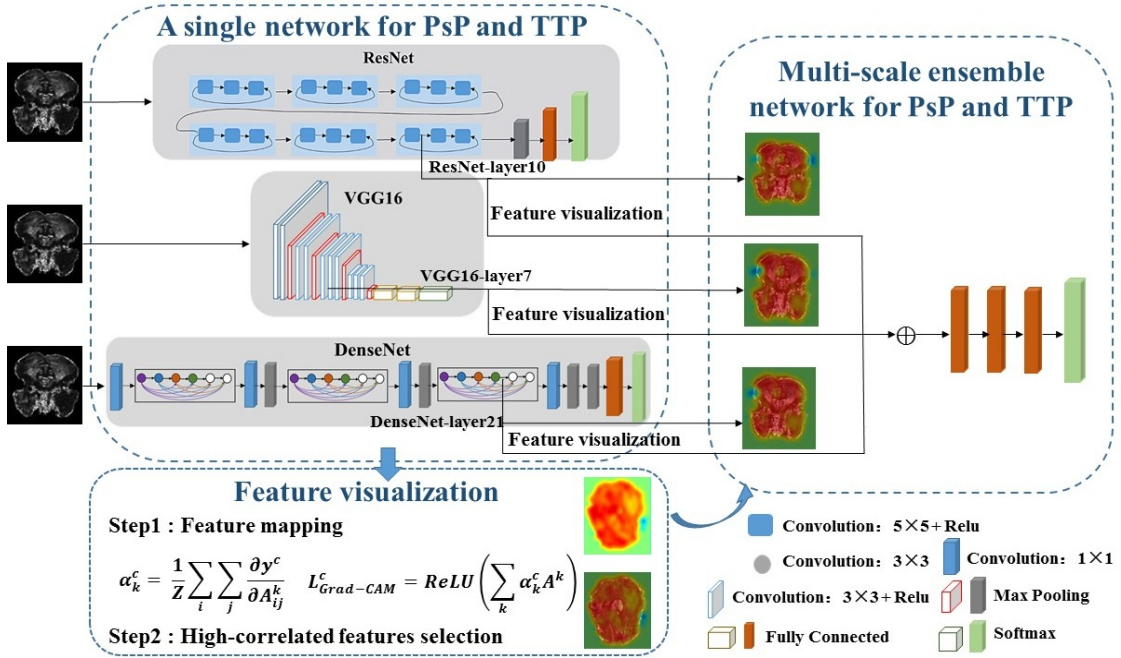

Fig.1. The Schematic for the interpretable and ensemble convolutional neural networks

## 2. Materials
### 2.A. Data collection

The data including clinical records and longitudinal DTI of 84 GBM patients (23 with PsP and 61 with TTP) were collected at the Wake Forest School of Medicine, the USA. DTI images were reconstructed in the matrix of 218×182 pixels. All of these patients received surgical resection and concurrent radiotherapy or chemotherapy with temozolomide subsequently. The enrolled patients received a similar dose (around 60 Gy) of conformal radiotherapy. Along with radiotherapy and chemotherapy, the patients underwent DTI examination (scanner: SIMCGEMR, GE Medical systems) every two or three months as monitoring. The final diagnosis of PsP and TTP was concluded form the follow-up images and professional evaluation of physicians.

2.B. Data augmentation

Since deep neural networks highly depend on the amount of data, augmentation is a generic way to improve CNNs. Data augmentation is effective to solve data insufficiency and reduce overfitting. Therefore, we apply a series of affine transformations to the valid images (total 1348 slices). First, all images are translated 5 pixels, 10 pixels, 15 pixels, 20 pixels and rotated by 3°, 6°, 9°, 12°, 15° respectively. Next, above 9 transformations and original images are rotated by 1°, 2°, 3°, 4°, 5° separately. So, the dataset was expanded dataset to 50 times its initial scale.

2.C. Data normalization

In order to eliminate the bias which is caused by different equipment, we applied the *Z-score* [29] method to normalize above dataset:

$$I' = \frac{I - \bar{I}}{\sigma_I} \tag{1}$$

where $I$ is a slice of dataset, $\bar{I}$ is the average grayscale value of dataset and $\sigma_I$ is the grayscale standard deviation of the dataset. After normalization, each slice in the dataset is transformed into the same magnitude, and it is suitable for further analysis and evaluation.

Finally, we randomly divide one tenth of the dataset as the testing set, and the other part is regarded as the training set.

3. Background works

3.A VGG for PsP and TTP

VGG [12] is the first design to explore the relationship between the architecture of CNN and its depth. The depth of VGG can be added steadily by using very small convolution filters. For all convolution layers (Conv.), the size of convolution kernel is fixed to 3×3, which is considered as the smallest size to capture pixel information from left, right, up, down and center. Spatial pooling is finished by max-pooling layers (MP) with 2×2 pixel window. Similarly, the main architecture is followed by fully connected layers (FC) and a softmax layer.

In this study, we applied a 17-layer VGG to diagnose PsP and TTP. The configuration of layers and corresponding channels are displayed in TABLE I.

TABLE I. Configuration of employed VGG

| Layer | Type | Channels | Layer | Type | Channels | Layer | Type | Channels |
|---|---|---|---|---|---|---|---|---|
| 1 | Conv. | 218×182×16 | 7 | Conv. | 54×45×64 | 13 | Conv. | 27×22×128 |
| 2 | Conv. | 218×182×16 | 8 | Conv. | 54×45×64 | 14 | MP | -- |
| 3 | MP | -- | 9 | Conv. | 54×45×64 | 15 | FC | 32 |
| 4 | Conv. | 109×91×32 | 10 | MP | -- | 16 | FC | 2 |
| 5 | Conv. | 109×91×32 | 11 | Conv. | 27×22×128 | 17 | Softmax | -- |
| 6 | MP | -- | 12 | Conv. | 27×22×128 | | | |

3.B ResNet for PsP and TTP

Although VGG realized a deeper network, some studies [26, 25] reported that there exposed some problems in such architecture. For example, the accuracy usually degrades rapidly after saturation with the increasing of depth. To solve this degradation problem, a deep residual learning framework which called ResNet is proposed. The residual learning block is shown in Fig. 2. For the input **x**, the block output **y** is defined as:

$$\mathbf{y} = \sigma(F(\mathbf{x}, W_i) + \mathbf{x}) \tag{2}$$

where the function $F(\mathbf{x}, W_i)$ is the *i*-th residual mapping; σ denotes relu [28] process.

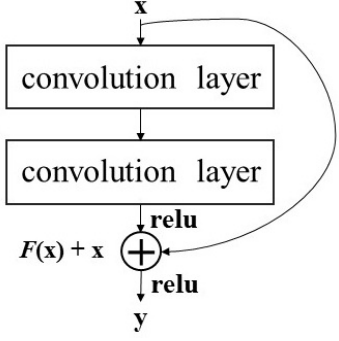

Fig.2. Residual learning block

We used a ResNet with six residual learning blocks to classify PsP and TTP. The kernel size of all the convolution layers is 5×5. After that, following a max-pooling layer, a fully connected layer and a softmax layer. The configuration of ResNet and corresponding channels are displayed in TABLE II.

TABLE II. Configuration of employed ResNet

| Block | Layer | Type | Channels | Block | Layer | Type | Channels |
|---|---|---|---|---|---|---|---|
| 1 | 1 | Conv. | 218×182×4 | 5 | 9 | Conv. | 14×12×64 |
|   | 2 | Conv. | 218×182×4 |   | 10 | Conv. | 14×12×64 |
| 2 | 3 | Conv. | 109×91×8 | 6 | 11 | Conv. | 14×12×64 |
|   | 4 | Conv. | 109×91×8 |   | 12 | Conv. | 14×12×64 |
| 3 | 5 | Conv. | 55×46×16 | -- | 13 | MP | -- |
|   | 6 | Conv. | 55×46×16 | -- | 14 | FC | 2 |
| 4 | 7 | Conv. | 28×23×32 | -- | 15 | Softmax | -- |
|   | 8 | Conv. | 28×23×32 |   |   |   |   |

3.C. DenseNet for PsP and TTP

Instead of exploring extreme deep networks, DenseNet [15] exploits the potential of neural network by feature reuse. The proposed dense block (Fig. 3) connects each layer to every other layer. That is to say, the feature maps of all layers are used as inputs for subsequent layers. Consequently, for each layer in DenseNet, the input of $i$-th layer is:

$$x_l = H_l(x_0, x_1, \ldots\ldots, x_{l-1}) \quad (3)$$

where $H_l(\cdot)$ is a composite function of batch normalization [29], relu [28] and convolution operation with 3×3 window. To facilitate down-sampling, dense blocks are usually connected by transition layers which are comprised of a 1×1 convolutional layer and a 2×2 average pooling layer.

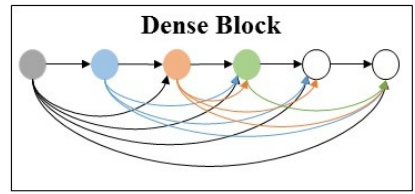

Fig.3. A dense block with six layers

We built a dense block with six convolution layers as Fig. 3. And we then set up a DenseNet with three dense blocks and two transition combinations to distinguish PsP and TTP. After that, following an average-pooling (AP) layer, a fully connected layer and a softmax layer. The configuration of DenseNet and corresponding channels are displayed in TABLE III.

TABLE II. Configuration of employed DenseNet

| Block | Layer | Type | Channels | Block | Layer | Type | Channels |
|---|---|---|---|---|---|---|---|
| Dense1 | 1 | Conv. | 218×182×8 | Dense2 | 13 | Conv. | 109×91×44 |
| Dense1 | 2 | Conv. | 218×182×12 | Dense2 | 14 | Conv. | 109×91×48 |
| Dense1 | 3 | Conv. | 218×182×16 | Transition2 | 15 | Conv. | 109×91×48 |
| Dense1 | 4 | Conv. | 218×182×20 | Transition2 | 16 | AP | 55×46×48 |
| Dense1 | 5 | Conv. | 218×182×24 | Dense3 | 17 | Conv. | 55×46×48 |
| Dense1 | 6 | Conv. | 218×182×28 | Dense3 | 18 | Conv. | 55×46×52 |
| Transition1 | 7 | Conv. | 218×182×28 | Dense3 | 19 | Conv. | 55×46×56 |
| Transition1 | 8 | AP | 109×91×28 | Dense3 | 20 | Conv. | 55×46×60 |
| Dense2 | 9 | Conv. | 109×91×28 | Dense3 | 21 | Conv. | 55×46×64 |
| Dense2 | 10 | Conv. | 109×91×32 | Dense3 | 22 | Conv. | 55×46×68 |
| Dense2 | 11 | Conv. | 109×91×36 | -- | 23 | FC | 68 |
| Dense2 | 12 | Conv. | 109×91×40 | -- | 24 | Softmax | 2 |

## 4. Proposed Interpretability guided ensemble network

### 4.A Feature visualization for PsP and TTP

Above three CNNs can realize stratification of PsP and TTP just on image-level label. That is to say, such networks have ability to distinguish subtle differences between brain lesion without region detection or segmentation. We have reason to believe that there are discriminative image regions to help CNNs identify the particular category. Therefore, feature visualization is a crucial step to raise the performance and reliability of deep networks.

Global average pooling (GAP) [30] is a vectorized strategy to replace fully connected layers in deep networks. GPA can effectively decrease redundant information and avoid overfitting in fully connected. More importantly, the feature maps can be easily converted into categories confidence maps by GPA. As shown in Fig.4, GAP takes the average value of each feature map for every convolutional layer instead of directly concatenating them as fully connected layers do.

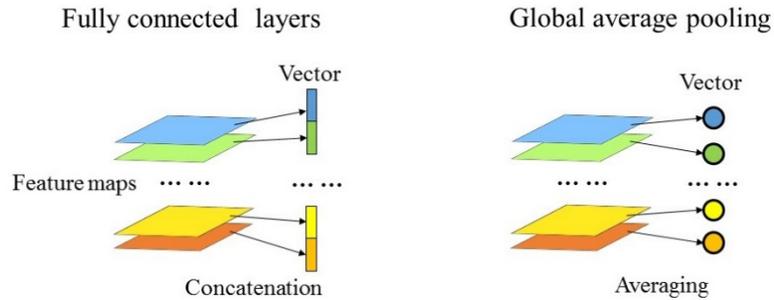

Fig.4. The comparison between fully connected and GAP

As illustrated in Fig.5, we perform GPA on all convolutional feature maps and use the new vectors to retrain softmax respectively, so the feature weights ($W$) for each class are acquired. Let the arbitrary convolutional layer produce $k$ feature maps $M^k$ of width $i$ and height $j$, a score $y^c$ for each class $c$ can be computed as:

$$y^c = \sum_k W_k^c \left[ \frac{1}{i \cdot j} \sum_i \sum_j M_{ij}^k \right]_{GAP} \tag{4}$$

Then the weight $\alpha_k^c$ is defined as:

$$\alpha_k^c = \frac{1}{i \cdot j} \sum_i \sum_j \frac{\partial y^c}{\partial M_{ij}^k} \tag{5}$$

This weight $\alpha_k^c$ represents a partial linearization of the feature map $M$ from the deep network, and captures the "importance" for the class $c$.

Next, the weight and feature map are combined and normalized by:
$$H = \text{Relu}(\sum_k \alpha_k^c \cdot M^k) \quad (6)$$

Finally, we map the obtained heat-maps, so-called gradient-weighted class activation mapping[31] (Grad-CAM) and realize feature visualization for all convolutional layers in the aforementioned CNNs. The computation is displayed in Algorithm 1.

---

**Algorithm 1: gradient-weighted class activation mapping for feature maps**

**Input:**  DTI slices $I$, feature maps $M$, feature weights $W$
$\quad\quad\quad$ $k$: the number of $M$, $c$: class, $N$: networks
**Output:**  Heatmaps $H$, Highlight images $L$
1: $\quad c \leftarrow 0$ (PsP) / 1 (TTP), $N \leftarrow 1$ (VGG)
2: **Repeat:**
3: $\quad$ **For** $c = 0$ to 1:
4: $\quad\quad y_N^c = ((W_k^c)_N, (M_k)_N)$
5: $\quad\quad {}_N^c = (y_N^c, (M_k)_N)$
6: $\quad\quad H_N = ({}_N^c, (M_k)_N)$
7: $\quad\quad L_N \leftarrow (I, H_N)$
8: $\quad N \leftarrow N+1$ ($N=2$:ResNet / $N=3$:DenseNet)
9: **Until** $N > 3$

Return: $L_N$

---

## 4.B Feature selection by experts

To gain a further understanding of deep features which responsible for category predictions, we subsequently design an expert-participated workflow to select the most relevant layers among the three CNNs.

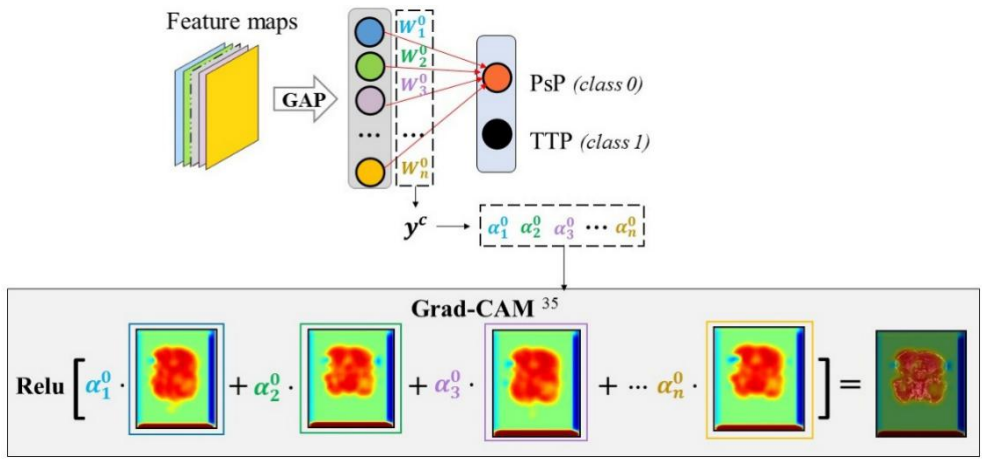

Fig.5. The flow chart of feature visualization

The expert group consists of 3 radiologists. They are all clinicians working at the Wake Forest School of Medicine, USA. In order to avoid empirical deviation, we deliberately choose the doctor with different levels. And their experience levels are detailed as follows: Radiologist1 is a chief physician; Radiologist2 is an attending doctor ; Radiologist3 is an attending doctor too.

As illustrated in Fig.6, the process of selection is divided into three steps. After scanning all the feature maps of convolutional layers, we find that different regions are activated by different layers. For example, some layers tend to fixate on high-density tissue, such as nasal septum, cella lateralis and middle cerebellar peduncle, while other layers focus on areas with relatively lower density, which may include lesion in DTI. So in the first step, the expert group is asked to disregard the layers which fail to locate the disease obviously. In this step, the layer will be abandoned as long as one radiologist reports irrelevance. Next, the radiologists are requested to respectively select the most relevant layers from the remaining layers. Finally, the final result is decided by discussion among the three candidate options.

We apply this workflow to all the convolutional layers of VGG, ResNet and DenseNet separately. And select the 7th layer in VGG, the 10th layer in ResNet and the 21st layer in DenseNet as the most relevant layer respectively.

4.C Multi-scale ensemble network

To take advantage of the expert knowledge and every CNN, we present an ensemble network to integrate the useful features. (Shown in Fig. 7).

As displayed in TABLE I, TABLE II and TABLE III, the number of the three groups of feature maps is 64. And the size of feature map in relevant layers are $54\times45$ of the 7th layer in VGG, $14\times12$ of the 10th layer in ResNet and $55\times46$ of the 21st layer in DenseNet. These feature maps are taken from different convolutional levels. In order to fuse multi-level information from different down-sampling scale, we generate a group of multi-scale feature maps.

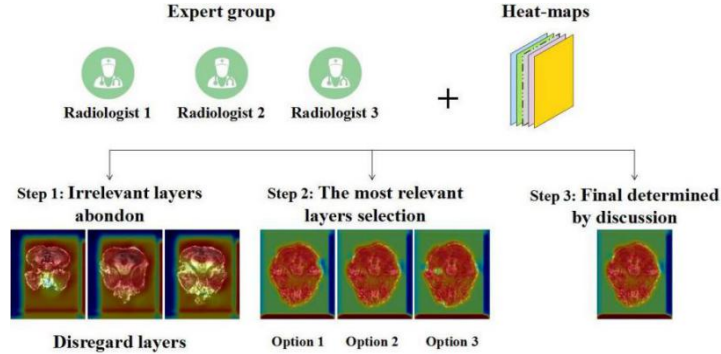

Fig.6. The workflow of relevant layer selection

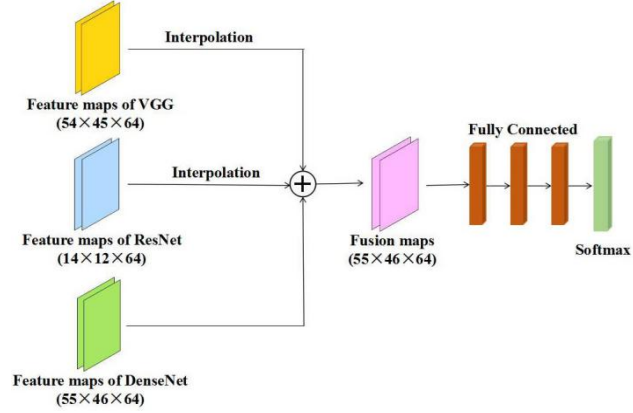
Fig.7. The framework of multi-scale ensemble network

To avoid information loss, we firstly use the bilinear interpolation to expand the small feature maps (14×12 and 54×45) to the same size as the biggest one (55×46). Then, to generate fused feature map, the value for the location *(i,j)* is computed as:

$$V_{i,j} = \frac{\left(V_{i,j}^{VGG} + V_{i,j}^{ResNet} + V_{i,j}^{DenseNet}\right)}{3} \tag{7}$$

Finally, we get 64 generations of multi-scale feature maps. And use them to re-train the network, which consists of three fully-connected layers and a softmax layer. The detailed sizes of inputs are shown in TABLE IV.

TABLE IV. Configuration of multi-scale ensemble network

| Layer | Type | Channels |
| --- | --- | --- |
| 1 | Input | 55×46×64 |
| 2 | FC | 64 |
| 3 | FC | 16 |
| 4 | FC | 2 |

5. Experiments and results
5.A. Experimental environment and settings

The complete experiments were performed on a workstation of Windows10 64-bit operating system with 16 GB memory and an NVIDIA GeForce GTX 1080 graphics card. The training and testing for VGG, ResNet, DenseNet and multi-scale ensemble network were developed on the DL library Keras with TensorFlow backend. The other algorithms were carried out on Python 3.5.

We separately train VGG, ResNet and DenseNet on the training set. Then feature visualization and selection are conducted. Finally, we use the selected feature maps to train the multi-scale ensemble network. In addition, the performances of each network are evaluated on the testing set.

In the training of VGG, optimization function is Adam with batch-size=50 and learning rate=0.001. In the training of ResNet, optimization function is Adam with batch-size=40 and learning rate=0.01. In the training of DenseNet, optimization function is Adam with batch-size=30 and learning rate=0.005. In the training of multi-scale ensemble network, stochastic gradient descent (SGD) is used as optimization function. The momentum of SGD is 0.9. The learning rate of 0.01 are considered training parameters.

5.B. The results

In this research, we realize the diagnosis of PsP and TTP with a total four deep architectures:

VGG, ResNet, DenseNet and the presented multi-scale ensemble networks. The accuracy of classification, sensitivity, specificity and the area under curve (AUC) are used to evaluate the performances of these networks. The classified results on testing set are displayed in TABLE V. And the receiver operating characteristic curves (ROC) are shown in Fig.8.

TABLE V. The results of four networks

| Network | Accuracy | Sensitivity | Specificity | AUC |
|---|---|---|---|---|
| VGG | 86.24% | 97.29% | 64.59% | 0.95 |
| ResNet | 87.68% | 99.44% | 64.63% | 0.98 |
| DenseNet | 88.02% | 99.24% | 66.04% | 0.98 |
| Ensemble Network | 90.20% | 91.26% | 88.18% | 0.99 |

For the single CNNs, TABLE V demonstrates that the accuracy gradually increases from top to bottom. Although DenseNet performs better than the other two networks, the enhancement is still limited. On the other hand, multi-scale ensemble network outperforms all CNNs and creates the highest accuracy (90.20%). Comparing with a single CNN, the ensemble network provides improvement with 3.96% for VGG, 2.52% for ResNet and 2.18% for DenseNet.

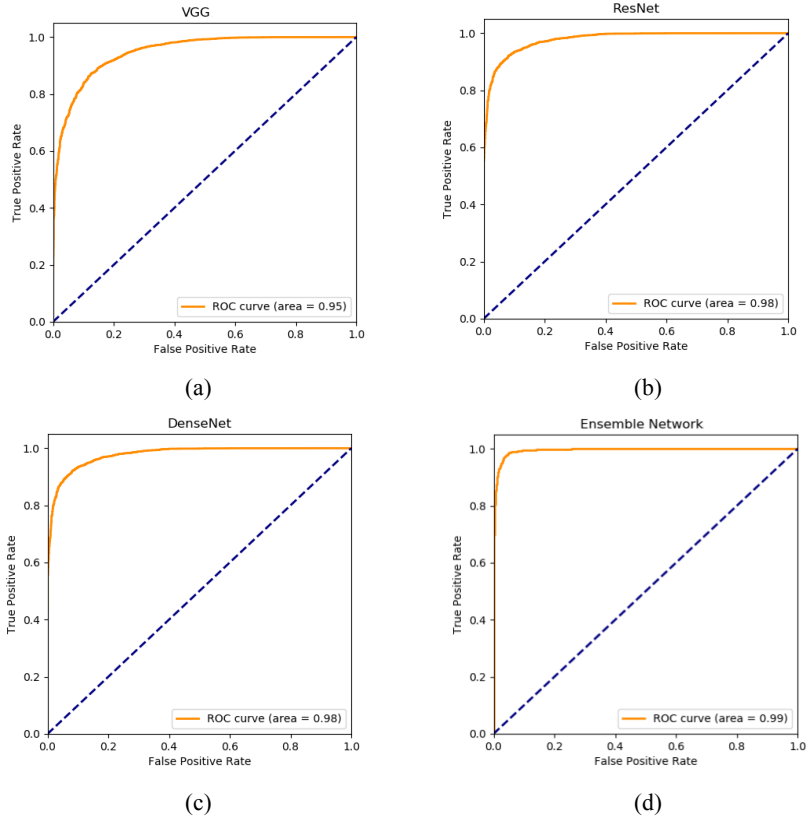

Fig.8. (a) The ROC of VGG; (b) The ROC of ResNet; (c) The ROC of DenseNet; (d) The ROC of ensemble network.

Although the sensitivity of the ensemble network is lower than those in other networks, its specificity reaches to 88.18%, which dramatically exceeds others. Besides, the gaps between sensitivity and specificity is all larger than 30% for VGG, ResNet and DenseNet. But such gap is only 3.08% for the presented network. In Fig.8, the ROC of ensemble network is the closest to top-left corner and it is not surprising that the ensemble network realizes the biggest AUC (0.99) among all the classifiers.

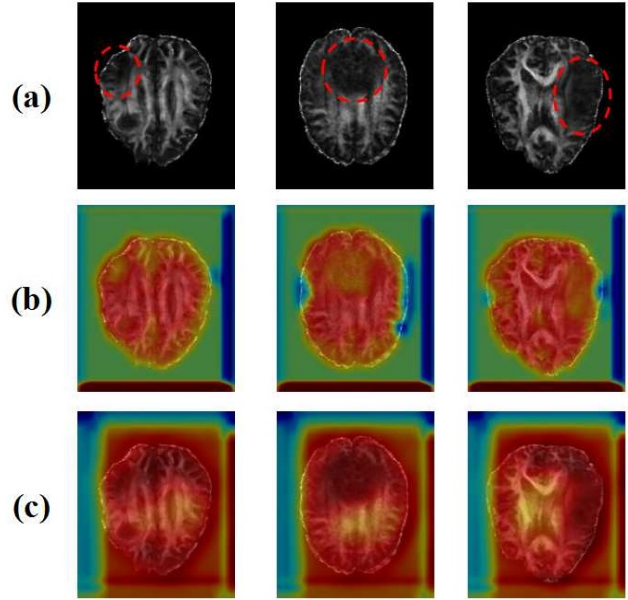

Fig.9. (a) The dotted line roughly outlines the lesion area; (b) Three examples of selected layers; (c) Three examples of excluded layers.

In addition, feature transparency can help us to explore the feature processing strategy of CNN. Some specific layers can effectively locate the suspicious regions even without segmentation. Since the related features are mixed with lots of unrelated features, the corporation with human knowledge will speed up feature compression and accurate feature selection. As shown in Fig.9, the selected layers (Fig.9(b)) cover lesion areas (outlined roughly in Fig.9(a)) approximately. In the meantime, the layers which focus on other regions (some examples are displayed in Fig.9(c)) are excluded.

TABLE VI. The number of features for four networks

| Network | VGG | ResNet | DenseNet | Ensemble Network |
|---|---|---|---|---|
| Number of features | 2,599,104 | 641,296 | 9,413,192 | 162,002 |

There are also fewer features in the presented network because of the transparency guidance. The results in TABLE VI illustrate that, compared to other CNNs, the ensemble network can use fewer features to realize higher accuracy.

6. Discussion

6.A The comparison of single CNNs and ensemble network

VGG, ResNet and DenseNet adopt different strategies to handle the deep features. They are currently regarded as the effective architectures of CNN. However, as TABLE V demonstrates, these typical CNNs realize just passable accuracy and show similar performances to each other. These sub-classifiers involve massive features, and the deep features are generated by various process including a series of up- and down- sampling in multiple scale. Although the single CNNs cannot satisfy diversity, the ensemble learning considers diversity by fusing the multi-scale features. In this way, the ensemble network enhances these single networks and brings a breakthrough. Among the improvements, the great enhancement of specificity is an important one. Since PsP is the mimicry of tumor progression at the tumor site or resection margins, it has a very similar appearance of TTP on DTI. Differentiation of PsP and TTP in GBM patients is still challenging even for experienced radiologists. In this context, accurately distinguishing PsP and TTP is a key but difficult step for GBM diagnosis. It is known that sensitivity is the rate of true positive while specificity is the rate of true negative rate. That is to say, higher specificity represents lower misdiagnosis and higher sensitivity represents lower underdiagnosis in medical problems. Therefore, high specificity is a significant requirement for the computer-aid diagnosis system of PsP and TTP. By multi-scale ensemble, the proposed structure combines the advantages of CNNs and achieves about 20% increase of specificity. The feature-based fusion optimizes the network and makes it more balanced and stable. Thus, it is promising for the diagnosis of PsP and TTP.

6.B. Visual Interpretability to guide the network construction

In contrast to the traditional medical image processing methods (such as clustering [32, 33]) where available knowledge can be treated as a set of prior rules, deep learning networks are totally black boxes between the inputs and outputs. And the characteristic of black boxes tends to be one of the largest disadvantages of deep neural networks. The ignorant of deep features hinders its theoretical improvements. For one thing, there must are some worthless information hidden in features. Excessive features will not only reduce the accuracy, but also make huge burden on hardware. For another, abstract features discourage human to trust in such architectures, especially in medical applications.

In this study, we use a gradient-weighted activation mapping method [31] to highlight important regions in all the feature maps of aforementioned CNNs. By such intuitive distribution of probability, radiologists are easy to recognize the disease-relevant layers. Different from mathematical methods, doctor's knowledge symbolizes human logic but is difficult to be described by the equation. We incorporate doctor's knowledge into the process of feature compression and selection and use it to guide the subsequent construction of an ensemble network. The results in TABLE V implies the feedback of clinicians and cooperation within human experience bring a breakthrough in the performance of an ensemble network. Besides, it confirms the significance of network understanding as well as the diagnostic value of tumor location in image-level classification.

7. Conclusion

In this paper, the authors put forward a multi-scale ensemble CNN to realize the automatic diagnosis of PsP and TTP on DTI. The network is novel with human knowledge-driven pattern and multi-scale feature fusion. As shown in TABLE V, the accuracy of presented network reaches to 90.20%. In addition, it achieves more than 20% promotion of specificity to the other CNNs. The results demonstrate the improvements can effectively optimize the performances and robustness comparing to single CNNs. So the presented network is a promising option for clinical diagnosis of PsP and TTP.

However, there are still some limitations in the research. As mentioned in section 2.A, the gold standard of diagnosis comes from the monitor for 2-3 months. In fact, longer follow-up may contribute more to automatic system construction. In addition, mathematical evaluation of experience is lacked in the process of feature selection. In the future, some quantitative metrics will be used to assess the effects of human knowledge.

8. Acknowledgements